# Combined Structural and Plasmonic Enhancement of Nanometer-Thin Film Photocatalysis for Solar-Driven Wastewater Treatment

*Desislava Daskalova[1,2], Gonzalo Aguila Flores[1], Ulrich Plachetka[1], Michael Möller[1], Julia Wolters[3], Thomas Wintgens[3], Max C. Lemme[1,2]\**

[1] AMO GmbH, Advanced Microelectronic Center Aachen, 52074 Aachen, Germany

[2] Chair of Electronic Devices, RWTH Aachen University, 52074 Aachen, Germany

[3] Institute of Environmental Engineering, RWTH Aachen University, 52074 Aachen, Germany

\*Email: lemme@amo.de / Phone: +49 2418867200

Titanium dioxide ($TiO_2$) thin films are commonly used as photocatalytic materials. Here, we enhance the photocatalytic activity of devices based on titanium dioxide ($TiO_2$) by combining nanostructured glass substrates with metallic plasmonic nanostructures. We achieve a three-fold increase of the catalyst's surface area through nanoscale, three-dimensional patterning of periodic, conical grids, which creates a broadband optical absorber. The addition of aluminum and gold activates the structures plasmonically and improves the optical absorption in the $TiO_2$ films to above 70% in the visible and NIR spectral range. We demonstrate the resulting enhancement of the photocatalytic activity with organic dye degradation tests under different light sources.

Furthermore, the pharmaceutical drug Carbamazepine, a common water pollutant, is reduced in aqueous solution by up to 48% in 360 minutes. Our approach is scalable and potentially enables future solar-drive wastewater treatment.

The use of titanium dioxide ($TiO_2$) nanoparticles or porous films is well established in heterogeneous photocatalysis. Under ultraviolet (UV) light irradiation, $TiO_2$ absorbs highly energetic photons, generating electron-hole pairs. These photoexcited charge carriers can then migrate to the surface of the semiconductor to participate in advanced oxidation processes. Owing to their large surface-to-volume ratio, $TiO_2$ nanoparticles in different modifications have been demonstrated to possess excellent photocatalytic performance at short wavelengths.[1] However, certain applications like wastewater treatment require stable $TiO_2$ films instead of nanoparticles, because of the inherent difficulties to filter and reuse the nanoparticles.[2] In addition, introducing large amounts of engineered nanomaterials into the biosphere poses toxicological and environmental risks.[3]

As an alternative, the advancement and wide availability of thin film technologies enables the application of thin film photocatalysis on an industrial scale. Nevertheless, thin film photocatalysts lack in efficiency compared to nanoparticles due to their lower surface area and fewer reactive sites. Hence, different strategies have been employed to improve their performance, such as thermally induced nanocrack networks in sputtered $TiO_2$ films to enhance the surface area[4], or semiconductor heterojunctions to enhance photocatalytic activity under UV[5,6] and extend the absorption to the visible range.[7] Utilizing plasmonic nanostructures to enhance photocatalysis is another promising route towards the commercialization of solar-driven photocatalysis.[8,9] It is based on the collective oscillations of electrons in the metal under illumination (plasmons), which generate intense and strongly localized electric fields in the vicinity of the metallic nanostructures. These plasmonic "hot spots", when coupled to a semiconducting photocatalyst, can then aid photocatalysis via effects like strong light absorption and scattering, electromagnetic field enhancement, and hot carrier generation. While its fundamentals are well understood, the interplay between different mechanisms of enhancement, like direct electron transfer and plasmonic resonance electron transfer at the metal-semiconductor junction, is still under discussion[10], in particular when moving away from the nanoparticle picture towards more complex nanostructure or multi-layer systems.[11] Designing novel devices and evaluating their performance therefore remains key to progress towards functional and sustainable solutions for advanced wastewater treatment.

Here, we present a combination of a periodically nanostructured broadband light-trapping three-dimensional surface, plasmonic metals like gold and aluminum (Au, Al), and a thin $TiO_2$ layer deposited by atomic layer deposition (ALD). Our approach avoids nanostructuring of noble metals, which is typically achieved by complex and time-consuming wet chemical processing. [12] Furthermore, similar to $TiO_2$ nanoparticles, the impact of such metal nanoparticles on living organisms and the environment remains unclear.[13–15] We circumvent these issues by incorporating the nanostructures in a glass ($SiO_2$) substrate, and then creating different material stacks with high surface areas on top of it. We demonstrate the efficiency of our photocatalytic devices through the reduction of methylene blue (MB) dye under different illumination wavelengths. Moreover, we show the successful degradation of the pharmaceutical pollutant carbamazepine under UV-A light using a nanostructured Au surface. Finally, we discuss our results with respect to the state of the art.

**Results and discussion**

We carried out photocatalysis experiments on five different sample types to evaluate the influence of the nanostructures alone versus their combination with the plasmonic metals (Fig. 1a, for process details see Methods section). The first reference sample (labeled "$SiO_2$" in figures) was a bare fused silica wafer cut to the standard experiment size, which was assessed to determine the effects of photolysis (if present). The flat surface sample with 25 nm $TiO_2$ was included again for investigating photolysis, but also standard photocatalysis of the $TiO_2$ itself (labeled "$TiO_2$"). The third reference sample, 25 nm $TiO_2$ on nanostructured $SiO_2$, is expected to utilize the effects of light trapping and surface area increase (labeled "Cones+$TiO_2$"). The last two samples differ in the choice of metal (Al, Au) and comprise of identically nanostructured $SiO_2$ covered with each metal, followed by the 25 nm $TiO_2$ ALD layer (labeled "Cones+Al+$TiO_2$" and "Cones+Au+$TiO_2$"). These samples combine the light trapping and surface area increase with the plasmonic effects in the metals and the Schottky barrier/electron transfer effects from the metal-semiconductor junction.

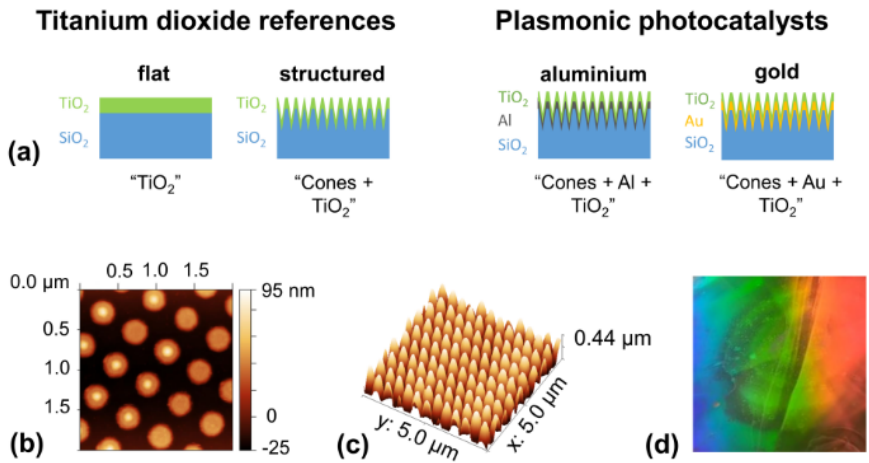

**Figure 1.** (a) Schematic of the final structures with active layer material combinations under investigation. (b) AFM image of the Al nanodisc hard mask created by nanoimprint lithography. (c) 3D AFM image of the conical nanostructure etched in SiO2. (d) Photograph of the Au/TiO$_2$ photocatalytic panel. The colorful surface effect is observed under ambient light at an angle by the naked eye or with a camera and is due to diffraction and thin film interference effects.

**Optical absorption and spectral overlap model**

The photocatalytic activity of the samples is expected to correlate with their optical absorption. Hence, the optical transmittance and reflectance spectra of the samples were measured by ultraviolet to visible light (UV-Vis) spectrophotometry (Perkin Elmer Lambda 1050) in the wavelength range from 250 to 800 nm. The samples were laid flat against the ports of a 150 mm integrating sphere housing the photodetector (photomultiplier tube). This ensured the collection of total transmittance (T) and total reflectance (R) for each sample, from which we calculated the total absorbance A = 1-T-R.

The UV-Vis spectra in Fig. 2a show that the flat TiO$_2$ reference has an absorption onset at 348 nm (dashed black line). It absorbs 55% of the UV-C radiation at 254 nm. The second reference sample with cones and TiO$_2$ (black line) shows a significant enhancement of UV absorption by 20% and an expansion of the absorption onset to 392 nm, which can be attributed to the light-trapping cones. There is an additional absorption peak of 26% at ~500 nm, matching the period of the structure. This is because the grid acts as a diffraction grating for light at a wavelength comparable to its period. Here, the subwavelength structure for light with wavelength larger than the period is turned into a partially diffracting grating for wavelength smaller than the period. These are known optical effects of gratings, which are less relevant for the photocatalytic application.[16]

The samples with Al (teal) and Au (orange) layers both show 95-97% absorption in the UV (250-320 nm) and over 70% across the visible and NIR spectrum (400-800 nm). Both types of metal/$TiO_2$ samples have low reflectance over the visible wavelength range, on average 11% for the Al sample and 3% for the Au sample (Fig. S3 in Supporting Information). In the case of gold, absorption reaches a maximum of 97 % at 500 nm; again, this effect is attributed to the grating period of the nanostructure. The Au-sample absorption spectrum is reproducible across different positions on the sample surface and different 5×5 cm² panels (Fig. S2). Our design was thus successful in creating highly efficient photocatalytic surfaces extended to a broader wavelength range. Fig. 2a further shows the source spectra of the UV-C and UV-A lamps we use in the reactor setup, with their peaks at 254 and 365 nm. This visualizes the spectral overlap between the light source driving the reaction and the absorption capabilities of the photocatalytic surfaces.

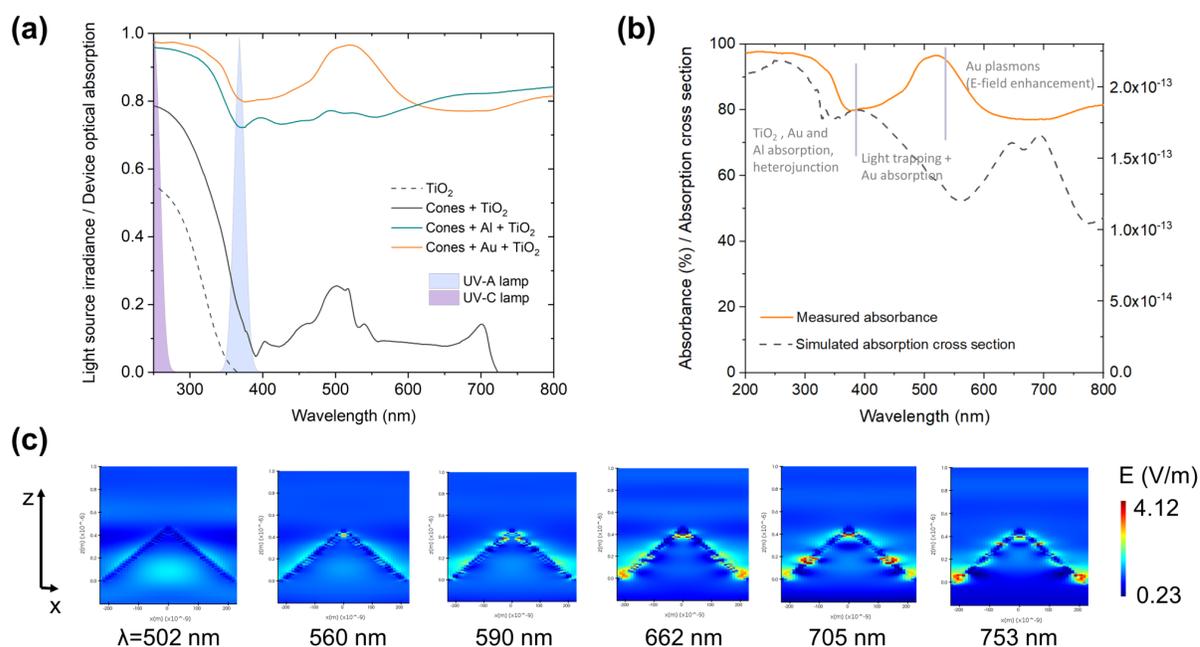

**Figure 2.** (a) Measured optical absorbance of the four fabricated samples superimposed with UV-C and UV-A light source spectra. (b) FDTD simulations of the $SiO_2$-Au-$TiO_2$ nanocone structure: absorption cross section in comparison to measured absorbance. (c) Simulated electric field cross section (*y*-plane) at different illumination wavelengths with plasmonic field enhancement ("hot spot" formation).

The photocatalytic efficiency can be estimated with an optical overlap model for plasmon resonance energy transfer (PRET)[17], which relates and quantifies the photocatalytic reaction with the electric field induced by surface plasmon resonance (SPR). According to the model, the photoactivity of a reaction is directly proportional to the overlap between the illumination source spectrum, the semiconductor absorption spectrum, and the metal SPR spectrum. Typically, nanometer-sized edges, holes or valleys in a metal surface can give rise to plasmonic "hot spots", or very intense electric fields in the vicinity of the metallic nanostructure. The energy is transferred non-radiatively from the plasmon to the semiconductor photocatalyst via the electric field in the near-field zone, generating electron-hole pairs in the photocatalyst.

We have simulated the optical properties of a periodic array of $SiO_2$-Au-$TiO_2$ nanocones on a $SiO_2$ substrate via finite-difference time domain (FDTD) solver of Maxwell's equations by Lumerical (for details, see Methods section). From this simulation, we have obtained the absorption cross-section for incident wavelengths between 200 and 800 nm, as well as the electric field distribution in the *xz* plane at different incident wavelengths. Fig. 2b shows the simulated absorption cross-section together with the measured absorbance and distinguishes between three spectral regimes. Namely, starting from the shortest wavelengths, the $TiO_2$, Al, and Au absorption regions extend to ~380 nm, where all three materials intensively absorb incoming photons. Then, between ~380 and ~520 nm only Au still has significant interband absorption, while the cone geometry still traps incoming light. Lastly, above ~520 nm and up until 800 nm, the dielectric properties of Au (the negative real part of the dielectric function and the imaginary part close to zero, see Fig. S7) can enable plasmonic field enhancement.

In addition, Fig. 2c shows cross-sectional images of the simulated electric field at illumination wavelengths from this last spectral region above 520 nm. This is where we expect to see the electric field hot spots. The simulations show that they are indeed present, localized on the cone tip for shorter wavelengths of 560 nm to 590 nm, then at the sidewalls for 662 nm to 705 nm. As the wavelengths are getting larger than the surface topography dimensions, the simulation shows the electric field building up between the sidewalls of neighboring cones and also in the trenches (753 nm).

Based on the spectral overlap between sample absorption and lamp irradiation, we predict the highest photocatalytic efficiency for the Au sample at UV-C and at UV-A, and possibly also for visible wavelengths due to the stronger extinction peak at 520 nm compared to the Al sample. The flat sample is expected to show the lowest efficiency.

**Methylene blue degradation**

We have performed dye degradation experiments under different illumination regimes corresponding to different regions of the samples' optical spectra in order to distinguish between the effects of photolysis, standard semiconductor photocatalysis, and the enhanced photocatalysis of the combined metallic and $TiO_2$ surfaces. The theoretical predictions based on the overlap model were experimentally evaluated using a self-built small-scale reactor in continuous flow mode (Fig. 3a, b).

The sample under study lies in a basin while an aqueous solution recirculates over it and a lamp shines light on it for the duration of the experiment. We chose three different lamps to study different regimes of the $TiO_2$ photocatalytic reaction: in-bandgap (UV-C) and out-of-bandgap (narrow band UV-A and broadband white light). The experiments with methylene blue (MB) were started with 30 minutes in the dark to allow for adsorption of MB on the sample surface. The MB degradation was monitored in real time by recording the reduction of its absorption peak at 663 nm with a red LED and a photodiode-based sensor (Fig. 3c). Every 10 seconds, the red LED shines through a well-defined volume of the MB solution and the photodiode measures the transmitted light. Hence, the photodiode signal is a measure of the water transparency, which corresponds to a decrease in MB concentration $C/C_0$, where C is the current concentration and $C_0$ is the initial concentration. The initial concentration of the MB solution was $10^{-5}$ mol $L^{-1}$ and the volume of the solution was 0.1 L. The transparency at this concentration level was measured and normalized to 1 in the figures, while a water transparency of 0 corresponds to clear water according to the detection limit of the sensor. The sensor was calibrated such that the light absorption intensity has a linear dependence on the MB concentration for the measurement range (see Supporting Information Fig. S5). The detection range was tuned with the sensor electronics for a large sensitivity between the initial and the final concentration. The water transparency level, related to $C/C_0$, should therefore always lie between one and zero. However, recordings of values larger than one are possible due to water evaporation, and do not indicate an increase in concentration beyond the initial $C_0$.

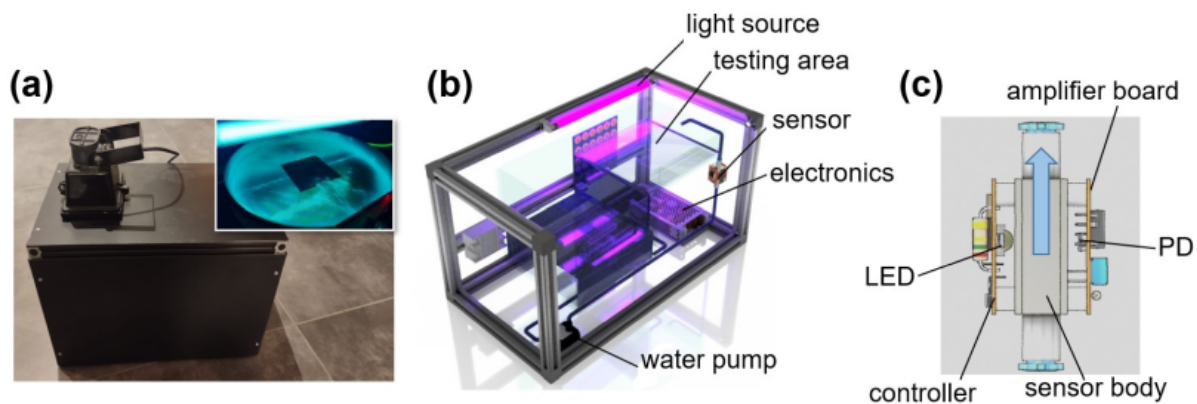

**Figure 3.** (a) Photocatalytic test reactor with lamp mount. Inset: basin with a 5×5 cm² sample and MB aqueous solution (0.1 L) under illumination. (b) 3D sketch of the reactor, comprising of light source over the testing area and water pump with piping to recirculate the liquid. (c) Close-up of the sensor system with red LED and photodiode (PD). The blue arrow indicates the direction of liquid flow through the well-defined volume of the sensor body.

First, we investigated the UV-C regime with a lamp peak at $\lambda = 254$ nm. The corresponding photon energies are larger than the bandgap of $TiO_2$ and there is high intrinsic absorption by the semiconductor. The real-time increase of water transparency under UV-C for all samples is shown in Figure 4. Methylene blue degradation, reactor data recorded in real time. (a) Under UV-C (254 nm) illumination. Dashed blue line: photolysis measured with a blank $SiO_2$ sample. Dashed black line: combined effect of photolysis and standard photocatalysis measured with a flat $TiO_2$ layer. Solid black line: combined effect of photolysis and standard photocatalysis measured with $TiO_2$ on a cone-structured surface. Enhanced photocatalysis is achieved by a stack of Al (teal line) or Au (orange line) and $TiO_2$ on cone-structured surfaces. The MB degradation measured on the reference fused silica substrate reflects photolysis (dashed blue line). After three hours of runtime, the final water transparency level due to photolysis only was 0.86. Similar results were obtained with the flat TiO$_2$ ALD layer (dashed black line) with water transparency reaching 0.84, indicating that the photocatalytic activity of the as-deposited flat thin film is negligible compared to the pure photolysis under UV-C. In contrast, the performance improved to a relative transparency level of 0.76 when using the sample with TiO$_2$ on the cone surface (solid black line in Fig. 4a). The reactor run with this sample combines the effects of photolysis and standard photocatalysis due to TiO$_2$ after the surface area enlargement. The increase in MB degradation is caused by the larger surface that can take part in the exchange of carriers with adsorbed dye molecules. This measurement then is a reference for the metallic samples, since structuring and layer thicknesses remain the same. As soon as one of the two metals are integrated on top of the glass nanostructure and are covered by the same TiO$_2$ layer, other enhancing effects come into play. The data of the two metal samples in Fig. 4a confirm the overlap model, as they show enhanced photocatalytic activity in UV-C compared to the TiO$_2$ on cones reference. The water relative transparency reached 0.51 and 0.37 for Al (teal line) and Au (orange line), respectively.

Second, we examined the UV-A regime with a lamp peak at $\lambda = 365$ nm, where there is no discernible photolysis and photon energies are below the bandgap of TiO$_2$. Hence, we expect no photocatalytic dye degradation unless there is a plasmonic enhancement by the metallic cone structures. The water transparency kinetics during a three-hour run are shown in Fig. 4b. The reference samples of flat TiO$_2$ (dashed black line) and TiO$_2$ on the cone surface (solid black line) remain at the initial MB concentration, which confirms the lack of photolysis and photocatalysis by TiO$_2$ alone. The substrates with Al (teal line) and Au (orange line), in contrast, increased the water transparency to 0.82 and 0.57, respectively. This is evidence for a plasmonically enhanced photocatalytic reaction.

We have fitted the data for the UV-A regime to first-order kinetics in Fig. 4c, using the calibration curve we had previously recorded (Fig. S5). The apparent rate constants obtained from the slopes are $k_{Au}$ = 0.02 min$^{-1}$ for the Cones + Au + TiO$_2$ sample and $k_{Al}$ = 0.005 min$^{-1}$ for the Cones + Al + TiO$_2$. For the Cones + TiO$_2$ sample (no metal), there is no apparent photocatalytic reaction, the concentration stays very close to the initial value and ln(C$_0$/C) stays very close to zero. Therefore, the apparent negative slope with $k_{TiO}$ = -0.002 min$^{-1}$ has no physical meaning. These results were later reproducibly confirmed on different days (example for the Cones + Au + TiO$_2$ sample on Fig. S3).

Third, MB degradation was investigated under a broadband white light source. In this case, only the Au sample showed photocatalytic activity, reaching a final water transparency level of 0.48 after three hours (Fig. 4d). The measurements taken from the other samples show an apparent increase in dye concentration because the lamp's excessive heat evaporated some of the water. This was observed despite a built-in Peltier cooling element, which was apparently insufficient. From Fig. 4d and the calibration curve (Fig. S5), we can estimate the influence of water evaporation to the dye concentration in the experiment with visible light. The final value of water transparency after three hours of white light illumination for the TiO$_2$ sample (no photocatalysis present) is 1.07, which is equivalent to a 1.8 V sensor signal. Given that at 2 V the MB concentration is 1×10$^{-5}$ mol L$^{-1}$ and the linear fit of our calibration curve, we estimate that 1.8 V corresponds to a 1.13×10$^{-5}$ mol L$^{-1}$ MB concentration. This is an increase in concentration of only one-tenth of the initial value and it leads to a water transparency decrease of 7% by the end of the measurement. Therefore, we consider the effect of water evaporation on the accuracy of the experiment to be within an acceptable margin, though modifications will be necessary for future experiments.

These three experiments serve to demonstrate the underlying mechanisms. In the UV-C case, the metallic nanostructures led to an improvement of efficiency by up to 51% at the three-hour mark for Au compared to conventional photocatalysis and photolysis. Here, the enhancement can be primarily attributed to the Schottky barrier at the metal-semiconductor interface, which lowers recombination losses of carriers generated by absorption in the $TiO_2$ layer. A linear extrapolation of the measurement data reveals that the metallic samples would clear the water volume completely 2.2 times (Al) to 2.8 times (Au) faster than the $TiO_2$ alone. In the UV-A case, the $TiO_2$ alone barely shows a photocatalytic reaction throughout the measured test interval. Here, the metallic nanostructures are required to induce photocatalytic dye degradation. In the white light case, Au nanostructures are required to induce catalysis.

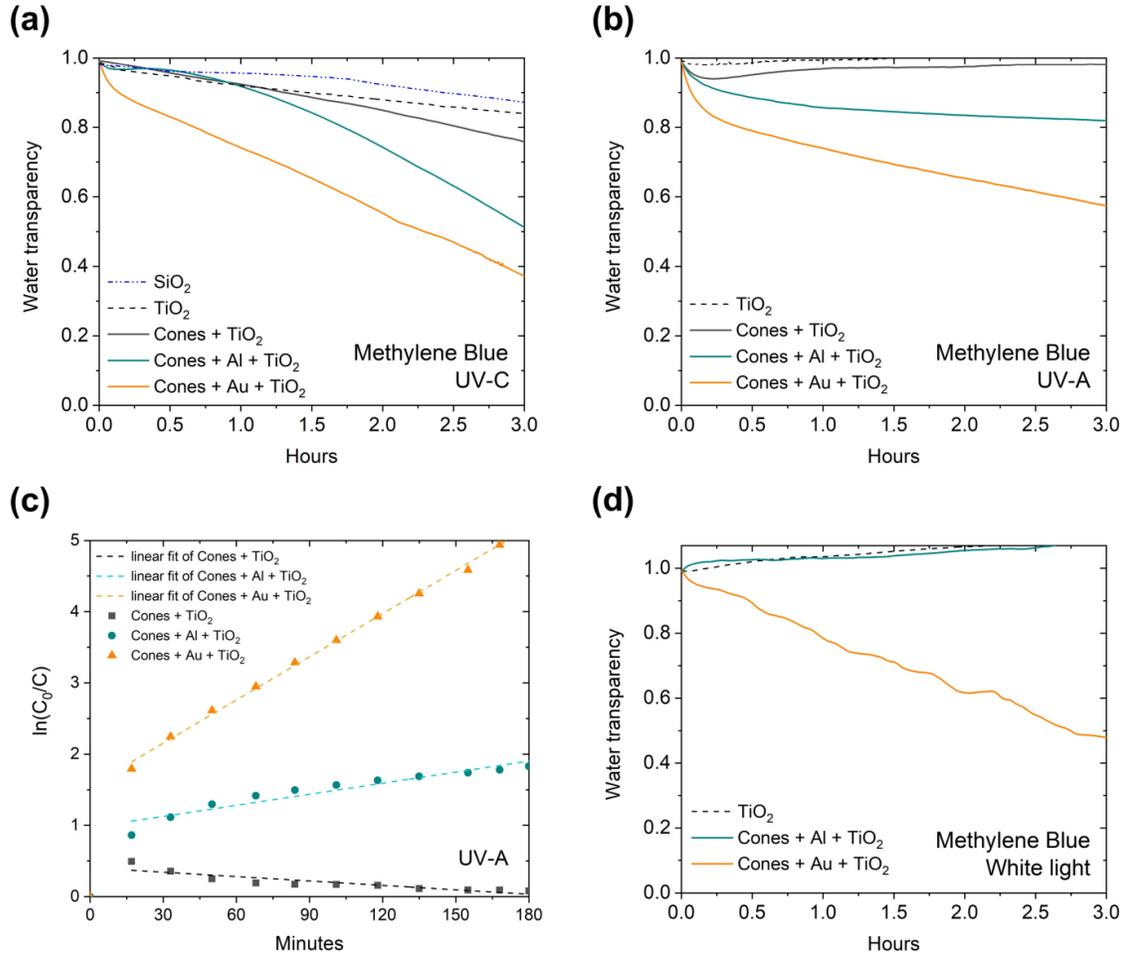

**Figure 4.** Methylene blue degradation, reactor data recorded in real time. (a) Under UV-C (254 nm) illumination. Dashed blue line: photolysis measured with a blank $SiO_2$ sample. Dashed black line: combined effect of photolysis and standard photocatalysis measured with a flat $TiO_2$ layer. Solid black line: combined effect of photolysis and standard photocatalysis measured with $TiO_2$ on a cone-structured surface. Enhanced photocatalysis is achieved by a stack of Al (teal line) or Au (orange line) and $TiO_2$ on cone-structured surfaces. (b) Under UV-A (365 nm) illumination. No photolysis or standard photocatalysis present, as measured with both a flat $TiO_2$ layer (dashed black line) and $TiO_2$ layer on the cone-structured surface (solid black line). Photocatalysis driven by plasmonics is demonstrated by the Al/$TiO_2$ (teal line) and Au/$TiO_2$ (orange line) layer stacks on cone-structured surfaces. (c) First-order kinetics fit of samples under UV-A illumination. (d) Under white light from high-pressure lamp. No photocatalytic reaction with a flat $TiO_2$ layer (dashed black line) or plasmonic photocatalyst with Al (teal line) present.

Lines show an apparent increase in dye concentration because of the lamp's excessive heating and subsequent water evaporation. The plasmonic photocatalyst with Au (orange line) successfully degrades the dye, the water transparency level increase due to evaporation is compensated for in this data set.

The superior performance of the Au sample compared to the Al sample at UV-C cannot be conclusively attributed to a difference in optical absorption at this wavelength (compare Fig. 2a). This indicates that other enhancement effects, like the Schottky barrier at the metal-semiconductor junction are driving the improved reaction in the in-gap regime. A higher Schottky barrier would promote better separation of the photo-excited charge carriers. Assuming the same electron affinity of $TiO_2$ in both cases, the Schottky barrier will be higher for the metal with the larger work function, as the barrier is given by the difference between the work function of the metal and the electron affinity of the semiconductor.[17] The work function of Au is typically in the range of 5.31-5.47 eV, while for Al it is 4.06-4.41 eV.[18] Therefore, the Au-$TiO_2$ interface would have the higher potential barrier and would be better at preventing electron-hole recombination.

The Schottky barrier plays an important role for enhancing photocatalytic performance by reducing recombination losses, if the attached semiconductor generates electron-hole pairs due to absorption. The curves measured under white light exposure (Fig. 4d) nonetheless show that neither the $TiO_2$, nor the Al-$TiO_2$ produce a decent measurable dye degradation. At wavelengths in the out-of-gap regime, the semiconductor does not absorb light and without generated carriers a sole reduction of recombination losses clearly cannot lead to any enhancement. Still, the Au-$TiO_2$ sample shows ample dye degradation. The carriers for this reaction presumably result from the interband absorption of light in the Au layer, followed by carrier transfer to the semiconductor. This transfer is most likely caused by a plasmonic enhancement in the nanopatterned metal. Here, the plasmonic field enhancement generates hot carriers in the metal that may overcome the Schottky barrier and be transferred to the semiconductor. The field enhancement is depicted in the simulation of the absorption cross sections in Fig. 2b. It is strongest between 620 nm and 750 nm where the imaginary part of the dielectric function of Au is lowest (Fig. S7). The E-field enhancement is visualized as hot-spots in the structure cross sections in Fig. 2c derived from the simulation data.

Benchmarking the performance of our photocatalytic surfaces to the state of the art in literature proves quite difficult because there is great variation in experimental parameters and conditions across reports – from sample composition, to irradiation wavelength and power, to test substance type, volume and concentration. Pedanekar *et al.*[19] recently published a table that collects the percentage of degradation efficiency in time as reported by different researchers. We summarize here only entries with MB as the model pollutant (Table 1) and conclude that our best-performing sample (60% in 180 minutes) matches well the state of the art for thin film photocatalysis. On closer inspection of the references, it becomes apparent that the cited percentages and times mostly refer to smaller testing volumes, for instance: 0.0035-0.005 L[20,21] or 0.03 L[22], 0.035 L[23], 0.05 L[24], compared to 0.1 L in our case. The initial MB concentration of $10^{-5}$ mol L$^{-1}$ in the literature references is the same as in our experiments. The only other reference with a testing volume of 0.1 L reports 60% degradation in 300 minutes.[25] Therefore, we conclude that our photocatalytic surfaces indeed constitute the state of the art in efficiency of purifying larger quantities of water.

**Table 1**. Photocatalytic performance of various metal oxide and sulfide thin films.

| Thin film photocatalyst | Synthesis method | Irradiation | Testing volume [L] * / Sample area [cm²] | Percentage of degradation efficiency in time | Ref. |
|---|---|---|---|---|---|

| N-doped TiO$_2$ | Sol-gel | Visible (150 W, Xe > 400 nm) | 0.0054.8 | 89% in 150 min | (20) |
|---|---|---|---|---|---|
| Cu-doped TiO$_2$/reduced graphene oxide | spray pyrolysis | UV-A, UV-B (300 W, 280-400 nm) | 0.0354 | 63% in 180 min | (23) |
| ZnO doped SiO$_2$ | Sol-gel | UV-A (20 W) | 0.11 | 60% in 300 min | (25) |
| ZnS | chemical bath deposition | UV–C (252 nm, 11 W) | 0.00352 | 92% in 240 min | (21) |
| Bi$_2$VO$_{5.5}$/Bi$_2$O$_3$ | chemical solution deposition | UV (300 W, Xe) | 0.0316 | 89.97%. in 300 min | (22) |
| CdS | SILAR | UV–visible (500 W, Xe) | 0.05 powder | 89% in 120 min | (24) |
| TiO$_2$-Au | magnetron sputtering/ALD | UV-A (365 nm, 6 W) Visible (125 W) | 0.125 | 60% in 180 min  50% in 180 min | This work |

*All entries are for methylene blue with concentration of $10^{-5}$ mol/L*

We further benchmarked our results with similar plasmonic photocatalysts and plotted the degradation time to reach maximum transparency (apparent C/C$_0$ = 0) in Fig. 5a. We included several reports based on sputtered TiO$_2$ films decorated with metal nanoparticles.[4,26,27] We also added results from metal-doped TiO$_2$ nanoparticle films fabricated by sol-gel methods[28–31] and TiO$_2$ layers without metals.[32] In the interest of fair comparison and to illustrate the complexity of making fair comparisons between vastly different approaches to photocatalysis and diversity in results reporting and presentation, we have also included (Fig. 5b) a table with the values of the apparent rate constant *k* reported in the studies referenced in Table 1 and Fig. 5a. Although the shortest times and highest values of *k* were achieved with TiO$_2$/metallic nanoparticles, there is still the open question of environmental exposure of nanoparticles. In addition, our data is in the same range as these nanoparticle-based methods and provides stable $TiO_2$ thin film surfaces, and may therefore be preferable overall. Looking beyond this initial performance evaluation, the possibility of iterative optimization of the nanostructure parameters and the choice of materials, together with the upscaling potential through large-area nanoimprint, makes the proposed photocatalytic surfaces an attractive concept for further studies.

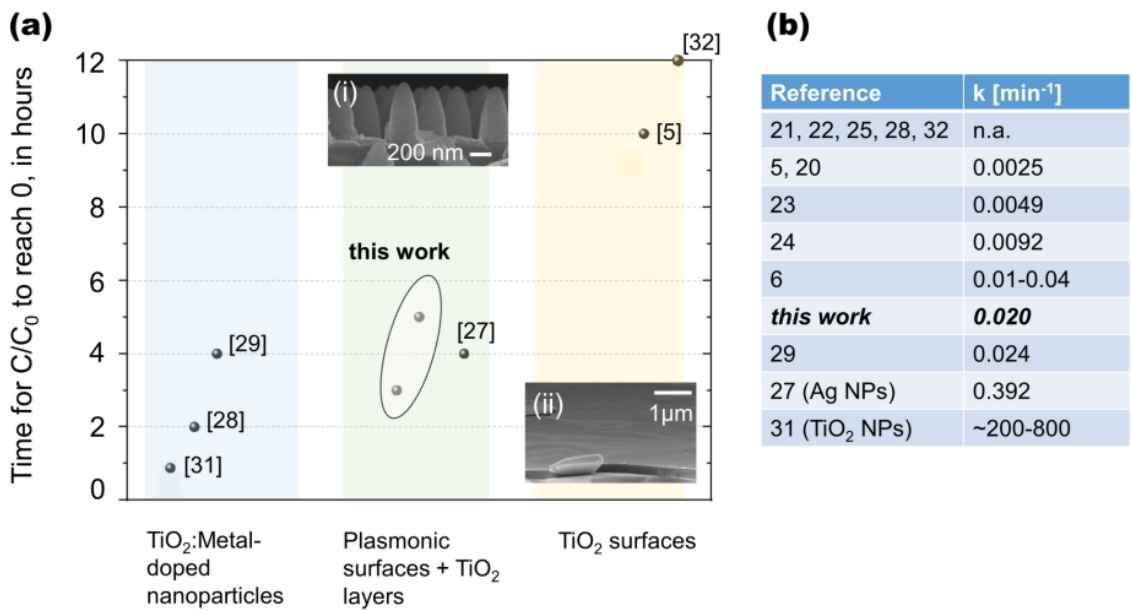

**Figure 5.** Comparison of typical MB degradation speed for different TiO2-based photocatalysts. (a) In terms of time for $C/C_0$ to reach apparent zero. Literature sources are indicated with references next to the points. Insets: SEM images of (i) gold-covered conical nanostructures as fabricated by us, (ii) a flat TiO2 ALD layer as deposited by us on silicon. (b) In terms of apparent rate constant *k* for different references.

**Carbamazepine degradation**

As a realistic real-life scenario, we investigated the degradation of the pollutant carbamazepine under UV-A illumination choosing the Au sample for its superior demonstrated performance. The sample surface, the reactor body and the tubing were thoroughly cleaned with distilled water to remove MB residues from previous experiments. Otherwise, the experimental system and the photocatalytic sample remained unmodified. However, carbamazepine has no absorption peaks in the visible spectrum. This prevented the real-time monitoring in the reactor with the red LED and photodiode combination. Instead, we measured the optical absorption spectra of the initial carbamazepine solution and the solution after 6 hours of processing inside the reactor *via* our UV-Vis spectrophotometer with 1 cm path length quartz cuvettes. The total solution volume in the experiment was 0.1 L and the initial carbamazepine concentration was 1 mg/L. For reference, carbamazepine detected in wastewater samples and in environmental waters is typically in the ng/L to µg/L range.[33] After six hours of reactor runtime, the characteristic carbamazepine absorption peak at 284 nm falls from 28.4% to 14.7% (Fig. 6). The second typical absorption peak close to 200 nm lies at the spectral limit of the measurement tool, therefore it is much harder to quantify. Nevertheless, the graph shows a clear reduction of the optical absorption — and hence of the concentration — of carbamazepine in the water, demonstrating the viability of our plasmonic nanostructure concept.

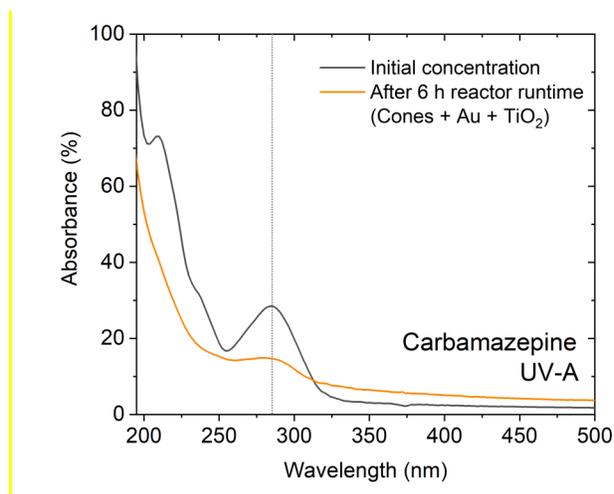

**Figure 6.** Carbamazepine degradation under UV-A illumination. Solution absorbance measured in UV-Vis spectrophotometer before and after reactor run time. Black line: absorbance spectrum of the solution with initial

concentration of 1 mg/L. Orange line: absorbance spectrum of the solution after 6 hours in the reactor with Au plasmonic photocatalyst. Characteristic peak at 284 nm reduced by 52%.

**Conclusions**

We have demonstrated enhanced photocatalysis efficiency *via* nano-structured and plasmonically engineered surfaces that act in synergy to circumvent the known limitations of thin film catalysts. Nanostructuring provides a three-fold surface area enlargement to increase the adsorption of reactants. In addition, the subwavelength dimensions of conical structures create a trapping effect for UV and visible light. This red-shifts the $TiO_2$ absorption onset to the violet range of visible light and enhances the UV absorption maximum to 78.4% for a $TiO_2$ thin film of merely 25 nm. Finally, a metal-semiconductor heterostructure formed on top of the textured surfaces (Al or Au and a $TiO_2$ ALD thin layer) further enhances the photocatalytic activity due to both promotion of charge carrier separation and plasmonic resonance. Specifically, the addition of Al or Au accelerates organic dye degradation under UV-C irradiation by a factor of 2.8 compared to bare $TiO_2$ and activates the photocatalytic degradation under UV-A, achieving up to 43% degradation in 180 minutes for methylene blue and 48% for carbamazepine in 360 minutes. Furthermore, the large plasmon resonance peak of the Au sample around 520 nm leads to methylene blue degradation of 52% in 180 minutes under broadband white light, which is relevant for solar-driven applications. We have achieved these results with substrates that were manufactured with scalable nanostructuring and conventional thin-film deposition methods on standard $SiO_2$ substrates. Furthermore, the approach is suitable for addressing other applications of photocatalysis, while being potentially less toxic than commonly employed nanoparticles. Our results hold promise for further up-scaling of thin-film based surfaces for wastewater treatment and other environmentally relevant sectors.

**Methods**

We first modelled the devices optical response *via* finite-difference time domain (FDTD) simulations with Lumerical FDTD photonic simulation software to predict a suitable nanostructure grating periodicity. The critical features for the gratings are their period / distance between structures and their depth of texturing (Fig. 1a). One prerequisite is that the grating period needs to be smaller than the wavelength of the incident light. Here, a period of less than or equal to ~500 nm is required for the enhancement of optical absorption in the UV and visible range. In addition, light-trapping is achieved when the height of the structures is equal to or larger than the wavelength of light, *i.e.* more than 300 nm in the case of UV-solar irradiation.

The chosen geometry closely matched the fabricated structure shape, including the dimensions of period and height of 480 nm, as well as thin film thicknesses of ~20 nm. Periodic boundary conditions were applied in the *x* and *y* direction, while a perfectly matched layer condition was set in the *z* direction. The simulation did not take into account surface roughness and effects arising from imperfections in the fabricated arrays. It was also limited to showing the response of the structures when the external electromagnetic field of the light interacts with them, which is not enough to explain the full photocatalytic reaction. Still, the simulation is a good first approximation to guide our understanding of the experimental results.

Common semiconductor technology tools and techniques can successfully produce structures that meet the theoretical requirements. Although this fabrication route may at first seem unusual in the context of $TiO_2$ photocatalysis, it has been a viable choice for other solar-driven devices like nanostructured solar cells.[34] Therefore, employing it for photocatalytic surfaces has the potential to enable innovations within the existing industrial infrastructure, as well as fabricating larger panels to cover the large surface areas required for wastewater treatment plants.

We used nanoimprint lithography (NIL) to fabricate nanostructures on 150 mm $SiO_2$ wafers (Fused Silica), which provides control of structural parameters like periodicity and lateral dimensions on the order of 10 nm.[35] From a single mould or master wafer, tens of nanostructured stamps can be readily produced, and each of them can be used to replicate up to 60 nanostructured wafers, corresponding to an area of ~1 m². Here, the master was made with laser interference lithography (LIL) in a setup operating with a laser wavelength of 266 nm. We exposed a 150 mm silicon wafer covered with UV resist with two expanded laser beams that interfered on the substrate (Fig. S1 in Supporting Information, process details in [36]). This created a periodic 2D grid of light and dark spots in the resist. The substrate was then rotated by 90° and a second exposure produced a pillar grating with pillar diameters of 240 nm, equal to half of the period defined as the distance from one pillar centre to the centre of an adjacent pillar (480 nm). Next, the resist grating was transferred into the silicon by reactive ion etching and the patterned silicon surface was coated with an anti-adhesion layer of octafluorocyclobutane. We then made a reusable soft stamp by casting polydimethylsiloxane (PDMS) onto the imprint master wafer.

The PDMS soft stamp was used to lithographically define the nanostructures of the fused silica wafers. However, the actual etching of the 450 nm deep conical structures in these wafers required an additional Al hard mask, which was deposited by DC magnetron sputtering. Next, AMONIL resist[37] was spin-coated, the soft stamp was pressed onto the resist, the resist was cured under UV illumination, and the stamp was removed. Afterwards, reactive ion etching with $BCl_3$-based plasma was used to transfer the pattern from the resist into the Al layer, forming Al nanodiscs as a hard mask on the $SiO_2$ surface (Fig. 1b). A second etching step with $CHF_3$-based plasma produced cones in the $SiO_2$ with an average height of 450 nm (Fig. 1c). The wafers were diced into 5×5 cm²-sized panels for further processing.

Either one of two metals (Au, Al) was deposited onto the cones *via* DC magnetron sputtering. The nominal thickness of the metals was 40-50 nm, *i.e.* if they had been deposited on a flat surface. The actual thickness over the conical topography can be roughly estimated by assuming a uniform distribution of the metal over the cones (SEM image in Fig. 5a). The geometry of the samples results in a threefold increase of the surface area compared to a flat surface. Thus, the average metal thickness is approximately 15 nm.

Finally, a TiO$_2$ layer with a thickness of t$_{TiO2}$ = 25 nm was deposited on top of the metals with a conformal ALD plasma process based on a titanium tetrachloride (TiCl$_4$) precursor at 300 °C. Fig. 1d is a photograph of the panel surface after nanostructuring and Au/TiO$_2$ thin film deposition. The structured glass acts as a surface relief diffraction grating for the incoming ambient light. This effect, combined with the thin film interference effect, results in a colorful pattern observable at an angle by the naked eye or with a camera. In addition to the Au/TiO$_2$ and Al/TiO$_2$ panels, a flat TiO$_2$ film on plain SiO$_2$ and a cone sample without metal but with TiO$_2$ were also fabricated as references, both with t$_{TiO2}$ = 25 nm, in addition to a reference fused silica wafer.

For the MB test in a lab-scale continuous flow photocatalytic reactor, our setup design and calibration were guided by the standard ISO 10678:2010[38] and consulting with the suggestions of Prof. Andrew Mills in his review of the ISO standard.[39]

**Corresponding Authors:**

\* Max C. Lemme – Chair of Electronic Devices, RWTH Aachen University, Otto-Blumenthal-Str. 2, 52074 Aachen, Germany; Email: max.lemme@eld.rwth-aachen.de

Ulrich Plachetka, AMO GmbH, Otto-Blumenthal-Str. 25, 52074 Aachen, Germany, email:

plachteka@amo.de

**Acknowledgement:** We acknowledge funding through the German Ministry of Education and Research, BMBF in the project PEPcat (02WCL 1519A) and the European Union's Horizon 2020 research and innovation program under the grant agreement No 101084261 (FreeHydroCells).

**Supporting Information Available:** Additional experimental details, including: a schematic of the laser interference lithography setup used to fabricate master wafers for nanoimprint lithography; Optical absorbance spectra of the nanostructured plasmonic photocatalyst with Au and $TiO_2$ measured on two different panels at three different positions each; Degradation of methylene blue by the nanostructured plasmonic photocatalyst with Au and $TiO_2$ under UV-A with reactor data from measurements on different days (PDF).

**Supporting Information**

# Combined Structural and Plasmonic Enhancement of Nanometer-Thin Film Photocatalysis for Solar-Driven Wastewater Treatment

Desislava Daskalova[1,2], Gonzalo Aguila Flores[1], Ulrich Plachetka[1], Michael Möller[1], Julia Wolters[3], Thomas Wintgens[3], Max C. Lemme[1,2]*

[1] AMO GmbH, Advanced Microelectronic Center Aachen, 52074 Aachen, Germany

[2] Chair of Electronic Devices, RWTH Aachen University, 52074 Aachen, Germany

[3] Institute of Environmental Engineering, RWTH Aachen University, 52074 Aachen, Germany

*Email: lemme@amo.de / Phone: +49 2418867200

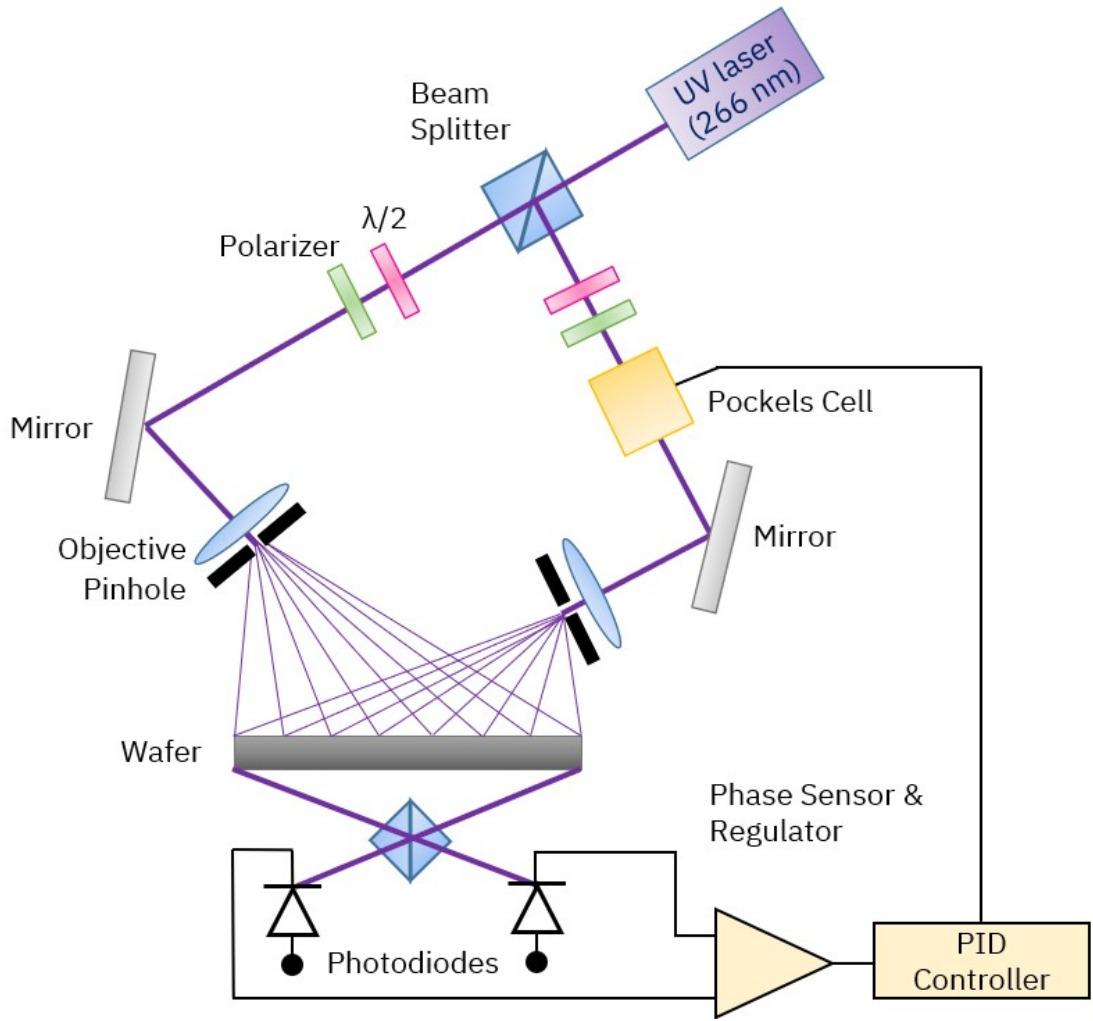

Figure S1. Laser interference lithography setup.

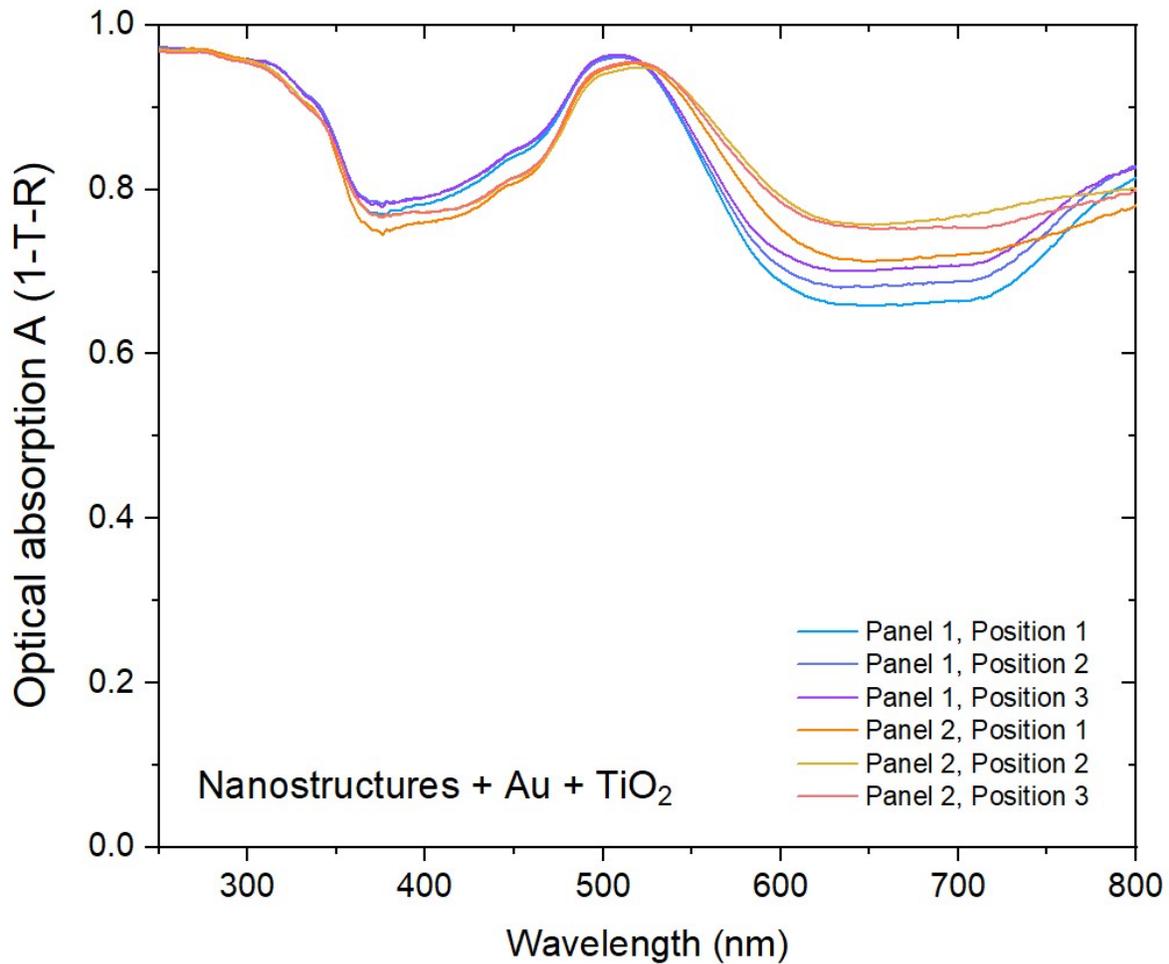

Figure S2. Optical absorbance spectra of nanostructured plasmonic photocatalyst with Au and TiO$_2$. Two different panels were measured at three different positions each. For Panel 1, the plasmon resonance peak is centred at 510 nm (A = 0.961), and for Panel 2 it is centred at 518 nm (A = 0.957).

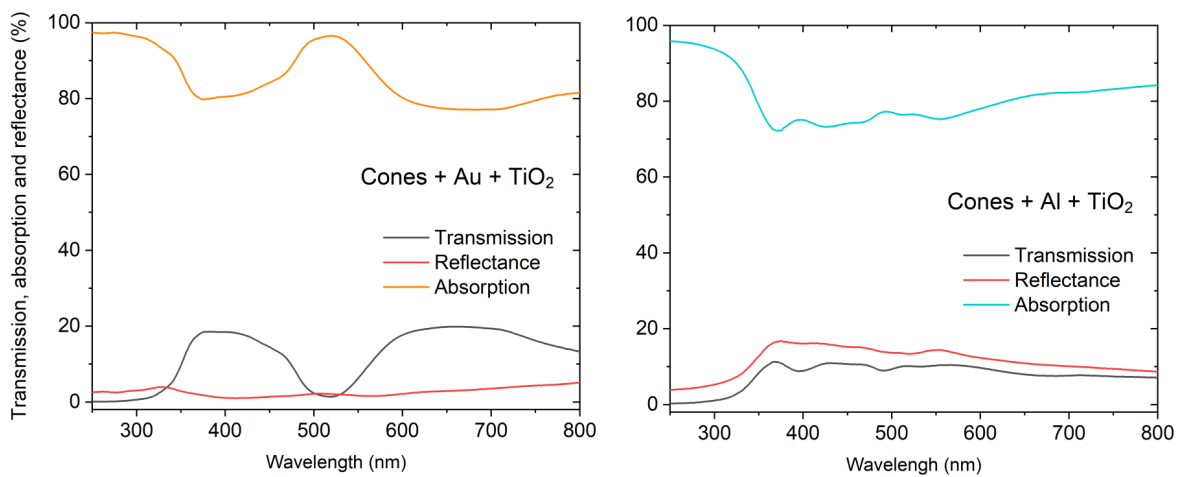

Figure S3. Full transmission (T), reflectance (R), and absorbance (A) spectra of nanostructured plasmonic photocatalyst with Au and $TiO_2$ or Al and $TiO_2$. T and R were measured and from them A was calculated as A = 100-R-T %.

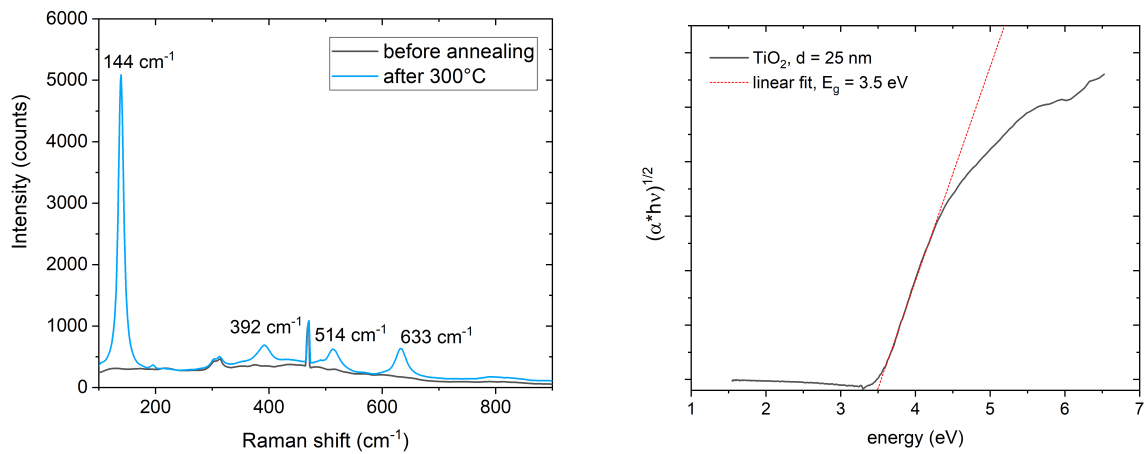

Figure S4. Left: Raman spectra of the pure ALD TiO$_2$ film as deposited and after annealing at 300 °C. Raman peaks characteristic to the anatase crystal phase can be identified after annealing. Right: Tauc plot of the ALD TiO$_2$ film with thickness of 25 nm calculated from the measured UV-VIS spectra. A linear fit indicates a bandgap of 3.5 eV.

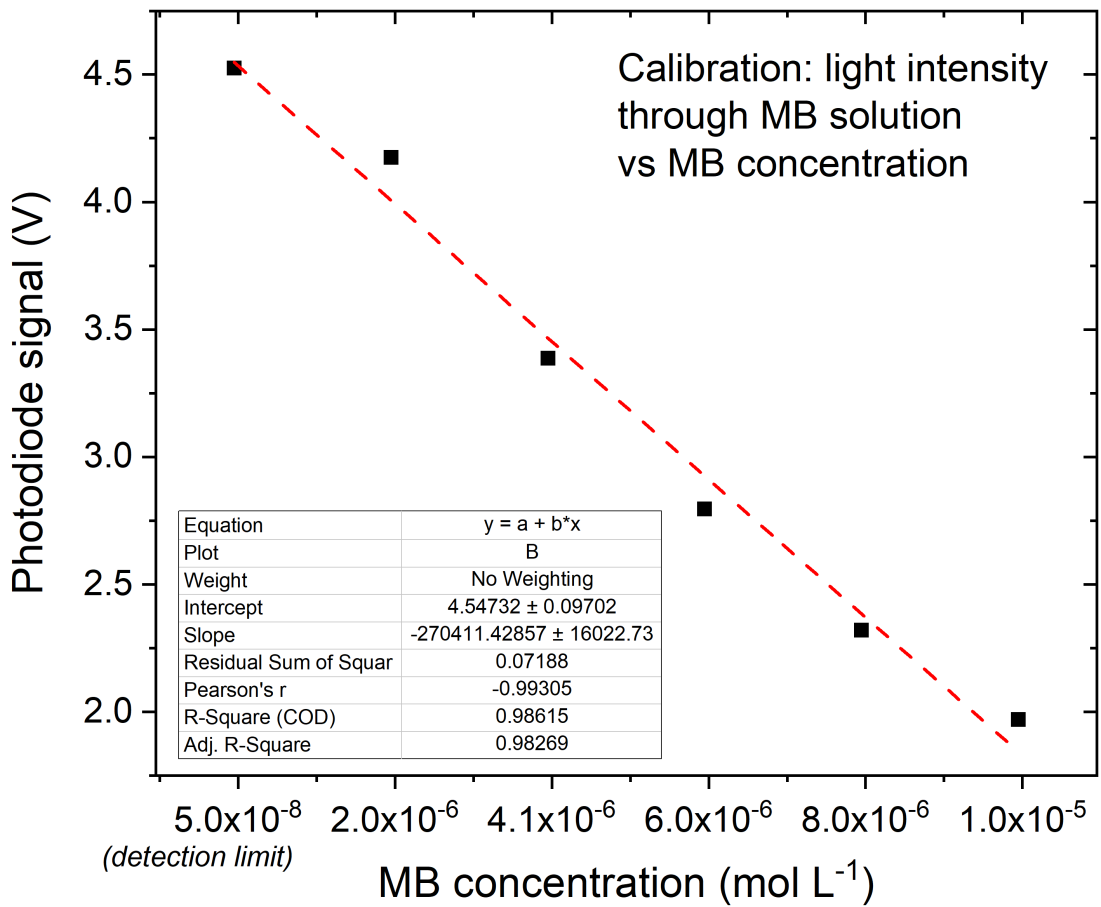

Figure S5. MB reactor sensor calibration for real-time monitoring of the water transparency. The light absorption intensity (as measured by photocurrent from the photodiode) has a linear dependence on MB concentration, with coefficient of determination R2 = 0.986. The circuit design of our photodiode-based sensor allows for the tuning the signal (photocurrent amplified and converted to voltage) to obtain a large sensitivity, i.e. large difference in signal level between $C_0$ and clear water: from 2 to 4.5 V. "Clear water" is defined by the detection limit of the sensor, where a further decrease in MB concentration would not lead to an increase in sensor signal beyond 4.5 V.

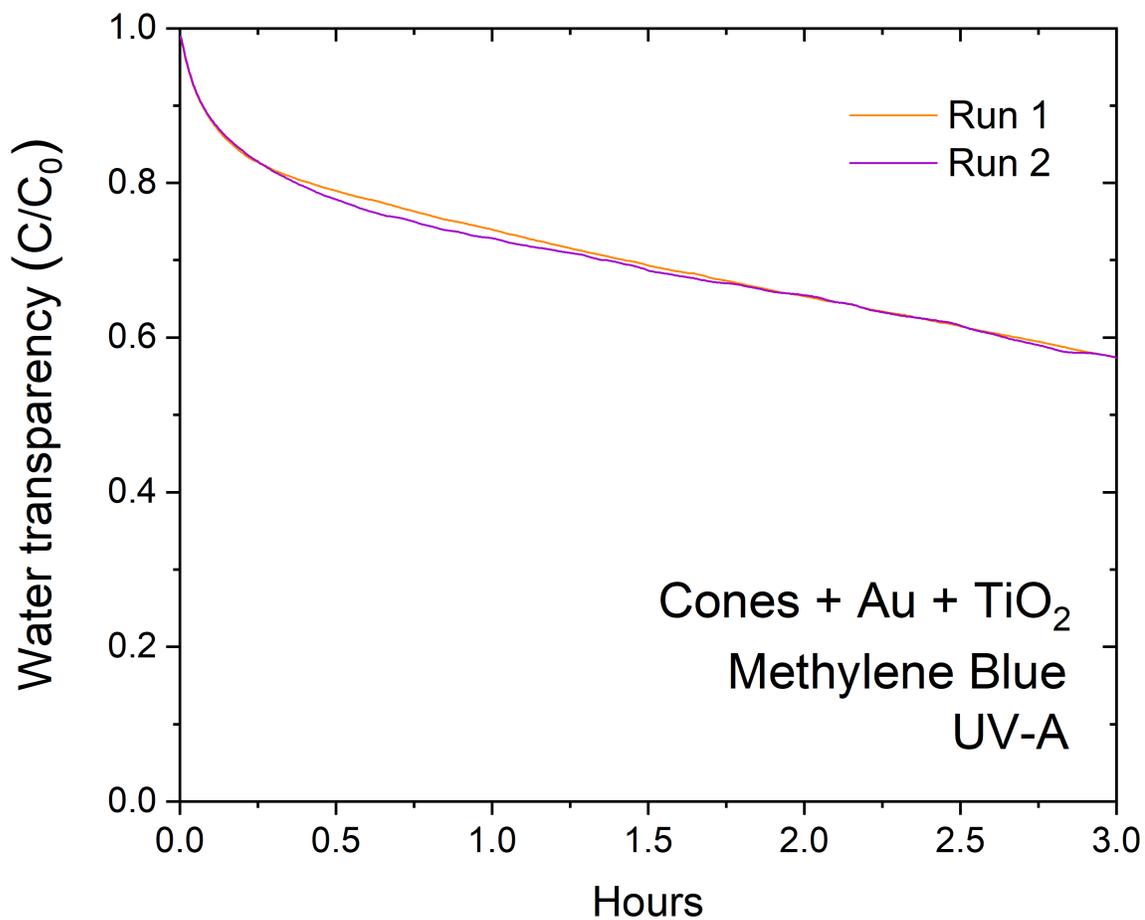

Figure S6. Methylene blue degradation, reactor data recorded in real time. Photocatalytic sample: Au/TiO$_2$ layer stack on cone-structured surfaces, under UV-A illumination. Two separate runs on different days (orange and purple line).

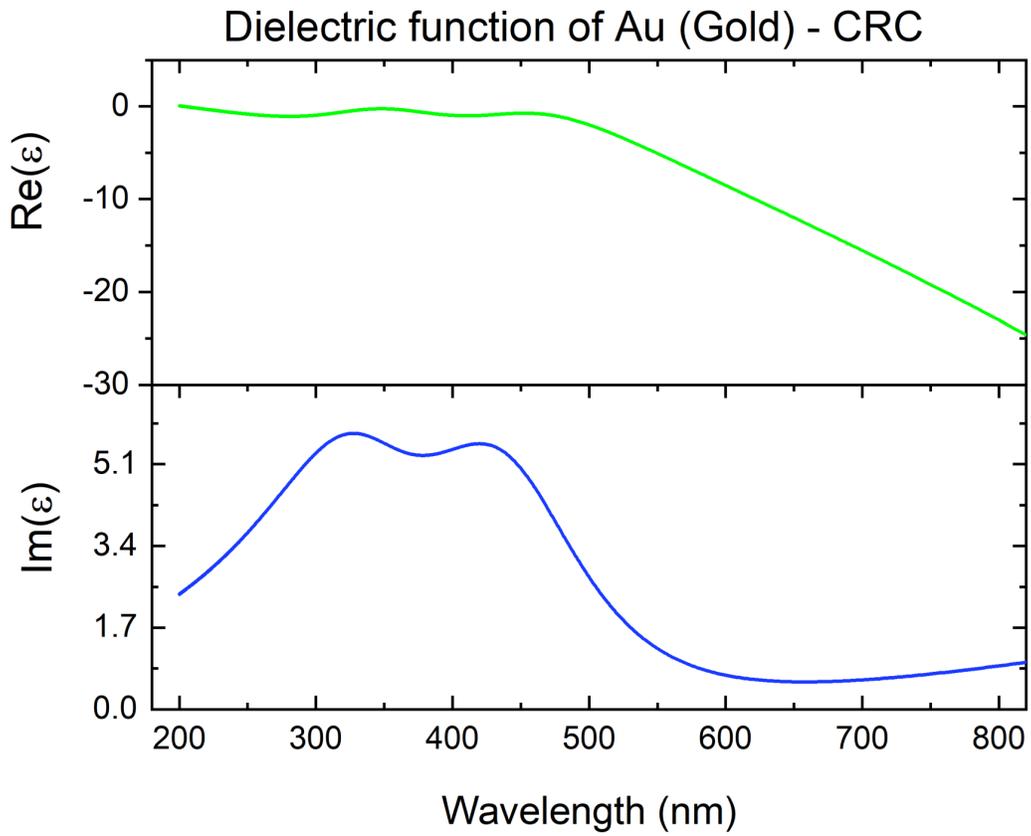

Figure S7. Dielectric function of gold, real and imaginary part: FDTD model of the CRC material data as provided in the Lumerical simulation software.